\documentclass[aps,prd,reprint,twocolumn,superscriptaddress,showpacs,notitlepage]{revtex4-2}
\usepackage{graphicx}
\usepackage{mathrsfs}
\usepackage{bm}
\usepackage{amsmath}
\usepackage{dcolumn}
\usepackage{epstopdf}
\usepackage{dsfont}
\usepackage{amssymb}
\usepackage{tabularx}
\usepackage{array}
\usepackage{float}
\usepackage{color}
\usepackage{epstopdf}
\usepackage{mathrsfs}
\usepackage[colorlinks, linkcolor=blue,anchorcolor=blue,citecolor=blue,urlcolor=blue]{hyperref}
\usepackage[T1]{fontenc}
\usepackage{hyperref}
\newcommand{\Rmnum}[1]{\uppercase\expandafter{\romannumeral #1\relax}}
\usepackage{graphicx}
\usepackage{chemformula} 
\usepackage[T1]{fontenc} 
\usepackage{easyReview}
\makeatother

\hypersetup{hidelinks,
	colorlinks=true,
	allcolors=blue,
	pdfstartview=Fit,
	breaklinks=true}

\begin{document}
\title{Ultrafast magneto-lattice dynamics in two-dimensional CrSBr driven by terahertz excitation}

\author{Yiqi Huo}
\affiliation{School of Physics, State Key Laboratory of Electronic Thin Films and Integrated Devices, University of Electronic Science and Technology of China, Chengdu 611731, China}

\author{Shuo Li}
\email{shuoli.phd@gmail.com}
\affiliation{Institute for Advanced Study, Chengdu University, Chengdu 610106, China}

\author{Luo Yan}
\affiliation{School of Mathematics and Physics, University of South China, Hengyang 421001, China}

\author{Ningbi Li}
\affiliation{School of Physics, State Key Laboratory of Electronic Thin Films and Integrated Devices, University of Electronic Science and Technology of China, Chengdu 611731, China}

\author{Sergei Tretiak}
\affiliation{Theoretical Division, Center for Nonlinear Studies, and Center for Integrated Nanotechnologies, Los Alamos National Laboratory, Los Alamos, New Mexico 87545, United States}

\author{Liujiang Zhou}
\email{ljzhou86@uestc.edu.cn}
\affiliation{School of Physics, State Key Laboratory of Electronic Thin Films and Integrated Devices, University of Electronic Science and Technology of China, Chengdu 611731, China}
\affiliation {Institute of Fundamental and Frontier Sciences, University of Electronic Sciences and Technology of China, Chengdu, 611731, China}

\begin{abstract}
Terahertz (THz) lasers provide a new research perspective for spin electronics applications due to their sub-picosecond time resolution and non-thermal ultrafast demagnetization, but the interaction between spin, charge and lattice dynamics remains unclear. This study investigates photoinduced ultrafast demagnetization in monolayer CrSBr, a two-dimensional material with strong spin-orbit and spin-lattice coupling, and resolves its demagnetization process. Two key stages are identified: the first, occurring within 20 fs, is characterized by rapid electron-driven demagnetization, where charge transfer and THz laser are strongly coupled. In the second stage, light-induced lattice vibrations coupled to spin dynamics lead to significant spin changes, with electron-phonon coupling playing a key role. Importantly, the role of various phonon vibration modes in the electron relaxation process was clearly determined, pointing out that the electronic relaxation of the  B$_{3g}^1$ phonon vibration mode occurs within 83 fs, which is less than the commonly believed 100 fs. Moreover, the influence of this coherent phonon on the demagnetization change is as high as 215 $\%$. These insights into multiscale magneto-structural coupling advance the understanding of nonequilibrium spin dynamics and provide guidelines for the design of light-controlled quantum devices, particularly in layered heterostructures for spintronics and quantum information technologies.

\end{abstract}
\maketitle

\section{Introduction}
Advances in nanoelectronics require smaller, more energy-efficient devices with faster storage and processing speeds \cite{jiang2019ultrafast, liu2020two, xiao2023review, song2023recent}. Potential solutions include using two-dimensional (2D) materials and encoding data in electron spin rather than charge \cite{takeda2022quantum, wang2024highly, islam2025advances}, as well as manipulating spins with ultrafast laser pulses\cite{anh2023ultrafast, emelianov2024ultrafast}. These pulses enable the study of ultrafast magnetization processes with minimal energy and high temporal resolution (femtosecond to picosecond timescales), allowing precise control of spin dynamics in real time \cite{hamamera2022polarisation, remy2023accelerating, davies2024phononic}. Terahertz laser pulses are located between far infrared and microwaves, and have the characteristics of sub-picosecond time resolution, wide frequency coverage, low photon energy, and strong penetration \cite{vicario2014gv, carnio2023intra, koulouklidis2020observation}. When Terahertz (THz) femtosecond pulses are applied, some interesting phenomena have been observed, including manipulation of collective spin excitations \cite{kampfrath2011coherent, wang2018magnetic, kovalev2018selective}, surface atomic adsorption \cite{huzayyin2014interaction, larue2015thz}, photoinduced transient birefringence \cite{baierl2016terahertz, zhao2020ultrafast}, and nonlinear electro-optical response \cite{mikhailov2007non, crassee2011giant, hafez2018extremely}. Compared with visible light excitation, the non-thermal, selective excitation and detection of THz lasers provide researchers with a new perspective for studying low-frequency spin electronics and quantum information science. 

Despite decades of research, the coupling interactions between lasers and material degrees of freedom in light-driven magnetization dynamics remain unclear, while spin dynamics involves complex interactions among spins, photons, electrons, orbitals, and phonons \cite{esposito2017nonlinear, nova2017effective, moroder2024phonon}. Laser pulses generate electromagnetic fields that interact with the spin and charge distribution of the material \cite{matsuzaki2015ultrafast, dang2020ultrafast, choi2024non}, while phonon-induced stress and deformation further influence spin dynamics and drive electron transitions between ground and excited states \cite{sharma2022making, cazorla2023giant, fonseca2024picosecond}. In the initial stages (< approximately 100 fs), spin-orbit coupling and torque dominate magnetic moment changes, while spin-phonon-mediated spin transfer between magnetic sublattices occurs on longer timescales (> approximately 100 fs) \cite{tengdin2020direct, he2023ultrafast}. During this phase, laser-induced lattice vibrations act as a reservoir for energy and momentum, absorbing angular momentum lost during demagnetization \cite{sharma2022making}. However, there is still uncertainty about how angular momentum is transferred from the spin system to the lattice phonon system (specifically, which phonon mode). To this end, we expect to reveal the microscopic mechanism after 100 fs of THz pulse excitation by studying the spin dynamics of two-dimensional materials with strong spin-lattice coupling. The recent discovery of layered van der Waals (vdW) magnets, such as the spin-dependent lattice anharmonicity of CrI$_3$  \cite{wang2021magnetic, pandey2021pivotal}, the light-tunable ferromagnetism of Fe$_3$GeTe$_2$ \cite{liu2020light, li2024optical} and light-controlled phonon vibrational modes of CrSBr \cite{torres2023probing, li20242d}, has sparked considerable interest in their spin dynamics in the 2D limit. Among these, the monolayer CrSBr is notable for its excellent air stability, semiconducting properties, and strong spin-phonon coupling \cite{bae2022exciton, pawbake2023raman}, making it an ideal platform for high-speed quantum information processing.

In this work, we provide a full ab initio description of the ultrafast THz photoinduced demagnetization dynamics in a monolayer material CrSBr, demonstrating its ability to rapidly demagnetize. Initially, electron-related demagnetization occurs within the first 20 fs, driven by charge transfer synchronized with the laser frequency. Without phonons, two other stages are observed, including thermal equilibrium and electronic rearrangement, with spin-orbit coupling (SOC) playing a key role. When electron-phonon coupling (EPC) is taken into account, electron rearrangement will occur at an exceedingly rapid rate. SOC and EPC provide distinct relaxation pathways, while specific single-phonon orbitals promote certain relaxation processes, and spin-flip transitions require multi-phonon and SOC interactions. The dual-pump pulse technique can be used to pre-excite a specific phonon vibration mode, and then a second laser pulse can be used to manipulate its spin dynamics. We pinpoint the complex relaxation process between various phonon modes and materials, and thus identify the B$_{3g}^1$ coherent phonon mode B $_{3g}^1$ that will greatly promote the relaxation process of electrons in monolayer CrSBr materials. This study elucidates the microscopic mechanisms of THz-induced magnetic moment changes, offering promising insights for quantum computing.

\section{Results and discussion}

The bulk of CrSBr exhibits a layered van der Waals (vdW) structure with the space group Pmmn (D2h). Its unit cell consists of two buckled CrS planes sandwiched between bromine layers, stacked along the c-axis (Figrue 1a).The structure under consideration appears to be rectangular, with an interlayer spacing of 7.96 Å. The a and b axis lattice constants are 3.50 and 4.76 Å, respectively\cite{esteras2022magnon, ziebel2024crsbr}. The crystal symmetry remains stable across a wide temperature range of 15–300 K \cite{telford2020layered}. The lattice constants a = 3.57 Å and b = 4.78 Å for the CrSBr monolayer have been optimized, which are in agreement with previous studies \cite{wang2023origin, yang2021triaxial}. 

\begin{figure*}[t!]
	\centering
	\includegraphics[width=0.8\textwidth]{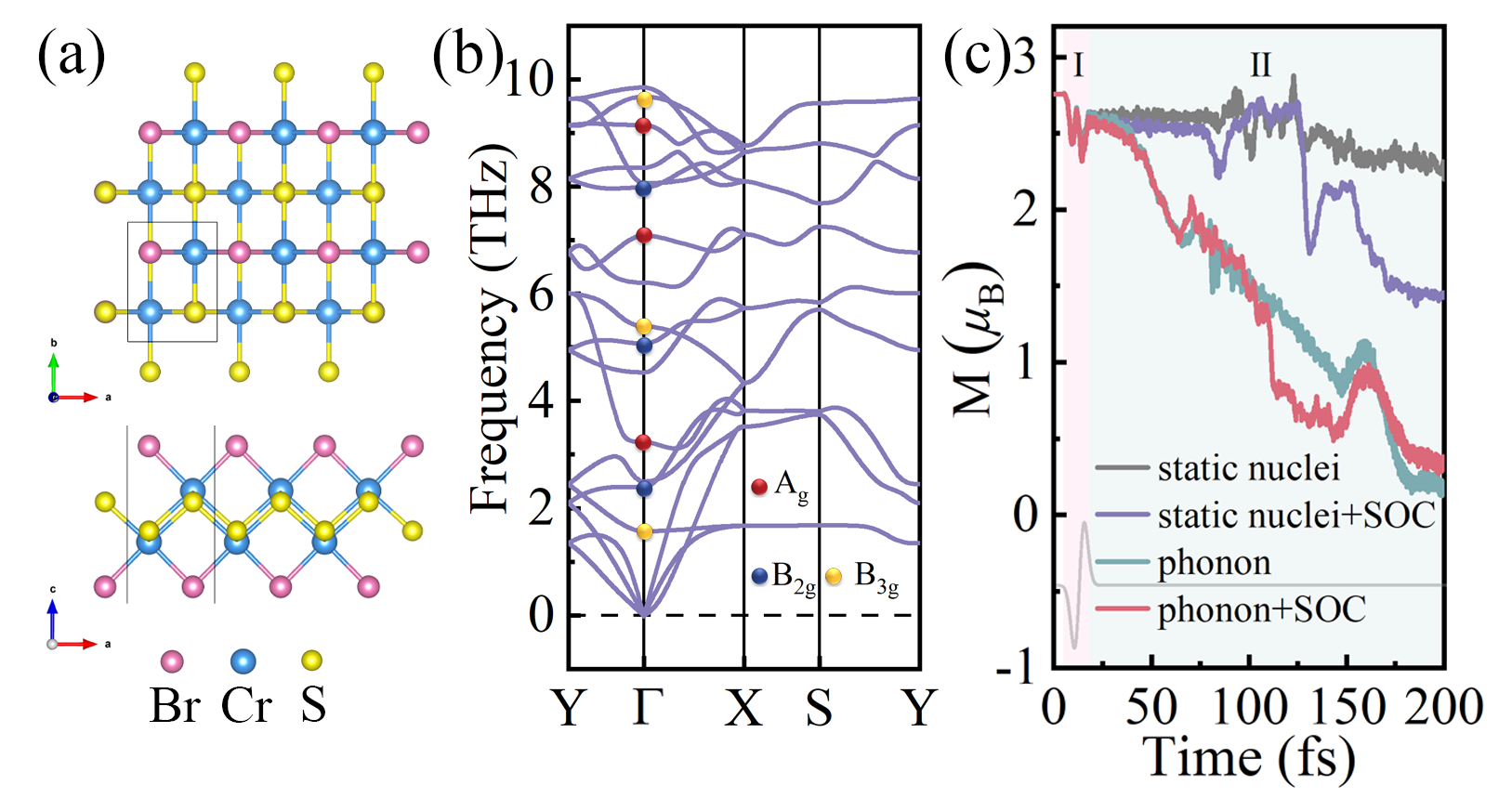}
	\caption{(a) Top view and side view of the atomic structure, (b) calculated phonon spectra, the phonon vibration modes in order from low to high frequency is marked as: B$_{3g}^1$, B$_{2g}^1$, A$_{g}^1$, B$_{2g}^2$, B$_{3g}^2$, A$_{g}^2$, B$_{2g}^3$, A$_{g}^3$ and B$_{2g}^3$). (c) Time-dependent magnetic moment dynamics of Cr$_1$ atom with respect to the situation of static nuclei, static nuclei+SOC, phonon and phonon+SOC. The profile of the laser pulse is represented by the grey line.}
\end{figure*}

we now discuss the vibrational properties of phonons in monolayer CrSBr. The air stability of the monolayer was confirmed through phonon spectroscopy, as shown in Figure 1b. In the first Brillouin zone, monolayer CrSBr exhibits 18 phonon modes, including 3 acoustic branches (with zero frequency at the $\Gamma$ point) and 15 optical branches, which Torres et al. confirmed this point \cite{torres2023probing}. At the $\Gamma$ point, these modes can be classified into the following irreducible representations:

$
$$\Gamma$$ = 3A$$_g$$ \bigotimes 2B$$_{1u}$$ \bigotimes 3B$$_{2g}$$ \bigotimes 2B$$_{2u}$$ \bigotimes 3B$$_{3g}$$ \bigotimes 2B$$_{3u}$$
$

Of the 15 phonon modes, A$_{g}$, B$_{2g}$, and B$_{3g}$ modes are Raman active, among which the A$_{g}$ mode can be detected in the first order Raman scattering, and the intensities of the other modes are relatively weak \cite{torres2023probing, jiang2024ultrafast, mosina2024electrochemical}. The vibrational frequencies corresponding to the verifiable Raman active modes can be found in Table S1.

For understanding the effect of phonon-spin interactions on the demonetization of CrSBr, the magnetization dynamics of monolayer CrSBr induced by ultrashort laser pulses were calculated based on real-time time-dependent density functional theory (rt-TDDFT) with Ehrenfest molecular dynamics. The laser pulse has a central frequency (\(\omega_0\)) of 10 THz, an incident pump energy of 9 mJ/cm\(^2\), a photon energy of 0.04 eV and a duration of 6.05 fs (the waveform and frequency distribution are shown in Figure 1c and S1a, respectively). In order to gain a clear understanding of the interactions between spins, electrons, orbitals and phonons in the spin dynamics process, we conducted a simulation of the laser-induced magnetization dynamics in monolayer CrSBr, including but not limited to one interaction. Therefore, four calculation schemes were established. The first scheme considered static nuclei, thereby eliminating the consideration of phonons and SOC; second scheme focused exclusively on SOC, while disregarding phonons; third scheme centered solely on phonons, while the fourth scheme incorporated both phonons and SOC. The phonons and SOC are employed to elucidate the atomic vibrations in the lattice and the coupling of spins and orbitals, respectively.

In accordance with the demagnetization curve of Cr$_1$ atom, as illustrated in Figure 1c (showing the z-direction magnetic moment when considering SOC), the magnetization dynamics process is subdivided into two demagnetization stages: \Rmnum{1} electron-related stimulated transition (0-18 fs); \Rmnum{2} relaxation of high-energy electrons (after 18 fs). In the absence of both phonons and SOC, the demagnetization amplitude is observed to be minimal (~0.4 $\mu $$_B$). Conversely, when phonons are taken into account, the amplitude reaches approximately 2.5 $\mu $$_B$. The following section will provide a comprehensive overview of each calculation method and the electron transfer mechanism at both stages.

As illustrated in Figure S2, the energy bands of a single-layer CrSBr unit cell exhibit a total band gap of up to 0.83 eV, while the ultrashort laser pulse photon energy employed for excitation is only 0.04 eV, but it can still excite electrons to a high-energy excited state. This is attributable to the potential for electron transitions in the opto-magnetic effect to occur via multiple pathways (Figure S1a). First, there is multiphoton absorption of at least two photons. When the laser energy is sufficiently high, electrons may absorb multiple photons simultaneously (Figure S1b), resulting in the total energy reaching or exceeding the band gap, which could potentially lead to the electron transitions. Second, interactions such as photoacoustic and optical-spin cause Raman scattering of electrons, resulting in changes of the photons frequency (Figure S1c). The final process is that of stimulated emission of radiation, whereby electrons return from high-energy excited states to the lower states (Figure S1d). We extend the work of Mrudul et al. and confirm that the stimulated motion of electrons in a material system is determined by multiple transformation processes \cite{mrudul2024ab}.

It can be observed in Figure 1c that the demagnetization happens temporal of each stage is nearly identical for all the calculation schemes. Given the slow oscillation of the Thz field, the electrons are able to discern the distorted electric field and transition to excited states before the laser field reverses, resulting in spin waves with a frequency aligned with that of the laser field. This stage signifies the inception of the ultrafast demagnetization process, a motion predominantly governed by the optically induced intersite spin transfer (OISTR) effect. Electrons drift away to high-energy hybrid and far-fold atomic nucleus, then tunnel to the non-nuclear attractors of the interstitial region. As the direction of the laser field changes, finally they return to the vicinity of the nucleus. The degree of demagnetization in this process is relatively modest (approximately 0.2 $\mu $$_B$), and the impact of phonon and spin-orbit coupling (SOC) is also limited. Therefore, the electron-related demagnetization process in the \Rmnum{1} stage is mainly electronic transitions dominated by multiphoton absorption and Raman scattering.

Moreover, there is another further decrease in the magnetic moment of the Cr atoms is observed. In the system without consideration of phonons, a period of thermal equilibrium will ensue, lasting up to 60 fs. In the stage \Rmnum{2}, the role of the SOC in static nuclei+SOC is of paramount importance, with an influence on the demagnetization amplitude reaching 0.8 $\mu $$_B$. Meanwhile, the findings demonstrate that phonons exert an influence on the redistribution of electrons in the material from the moment (18 fs) the pulse intensity has not yet reached zero. This phenomenon is independent of the spin-orbit coupling (SOC). Therefore, the SOC switch will be always on in our study. The magnetic moment of Cr atoms is observed to be almost 0$\mu $$_B$ at the conclusion of the simulation. These point to the conclusion that the second demagnetization is a process that is regulated by a variety of mechanisms, which include nonlinear and higher-order interactions \cite{mrudul2024ab}. A detailed analysis of the microscopic mechanism of the second demagnetization process will be conducted for each calculation scheme.

With the presence of SOC, we directly observe a substantial magnetic moment change start after 82 fs (Figure 1c). It is evident that SOC plays a significant role in the study of spin dynamics of magnetic materials. Accordingly, the nonlinear magnetic moment was calculated and observed to undergo a change over time, see Figure 2a. Subsequent to 82 fs, the magnetic moment of Cr$_1$ atoms underwent multiple demagnetization, subsequently recovering to varying extents following the first two demagnetization. However, at 125 fs, an observation was made that the spin shifted from out-of-plane to in-plane orientation, with the in-plane magnetic moment exhibiting vibrations around 0 $\mu $$_B$, reaching a peak value of 0.43 $\mu $$_B$. While the out-of-plane magnetic moment exhibits a rapid decay, declining from a maximum value of 2.68 $\mu $$_B$ to 1.72 $\mu $$_B$ within [126-132] fs. It is anticipated that the thermal equilibrium electrons will undergo a rearrangement and then experience vibration in order to attain stability.

\begin{figure*}[t!]
	\centering
	\includegraphics[width=0.8\textwidth]{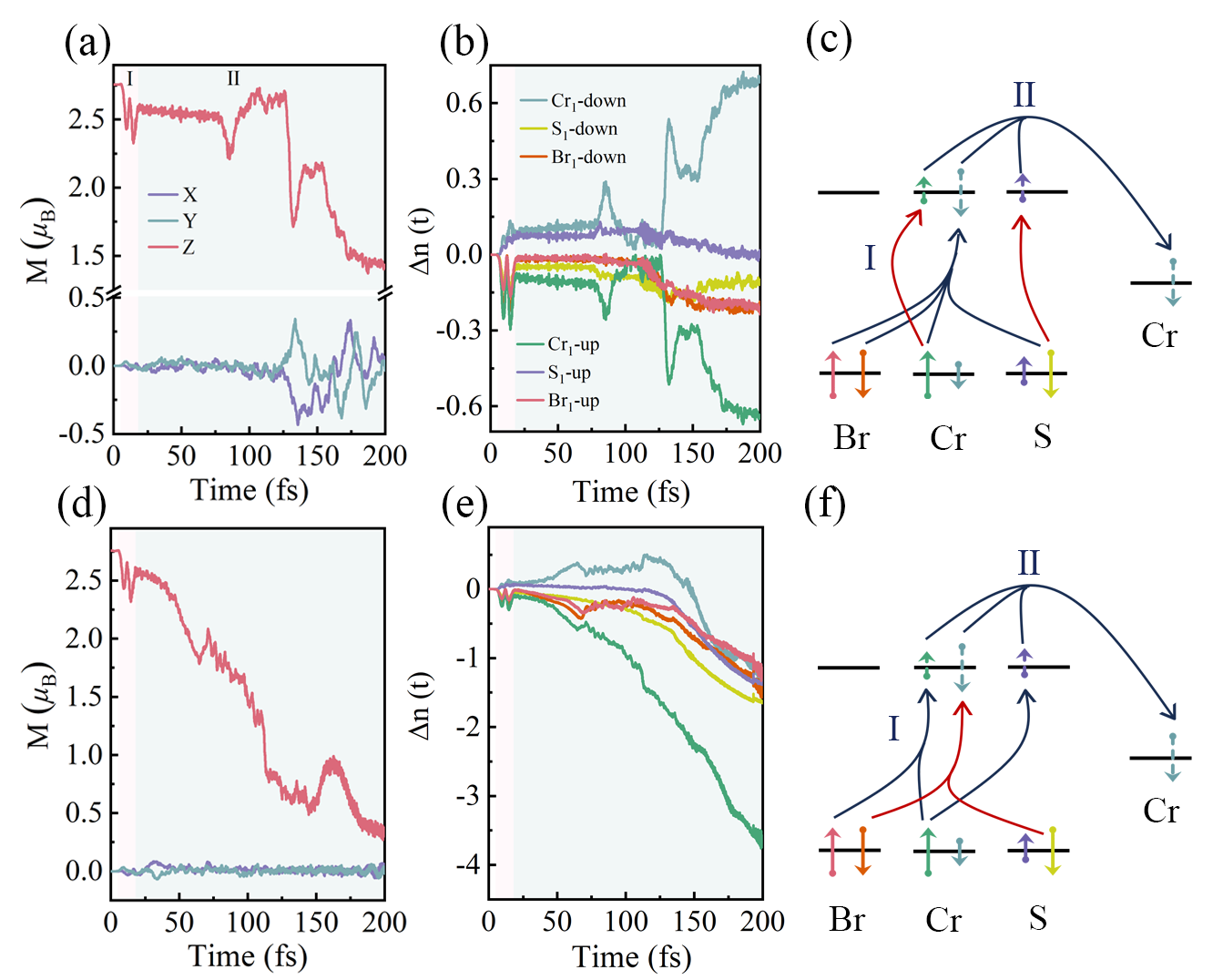} 
	\caption{Light-induced time-dependent dynamics in CrSBr monolayer (a)-(c) with static nuclei+SOC and (d)-(f) with phonon+SOC. (a), (d) The magnetic moment for Cr$_1$ atom in the different directions. (b), (e) The change in the spin-resolved charge ($\Delta$n($\emph t$) = n($\emph t$) - n(0)) evolution. (c), (f) Schematic diagram of spin-resolved charge transfer.The straight arrows placed on the horizontal lines are indicative of electron occupancy. In addition, the solid and dotted arrows are used to symbolize valence band and excited state electrons, respectively. The longer the arrow, the larger of occupancy. The color of the arrows aligns with (b) and (e). The curved arrows denote two demagnetization stages. }
\end{figure*}

To gain further insight into the underlying mechanism of the observed change in magnetic moment with SOC, we conducted a detailed investigation into the time-dependent spin polarization dynamics of the electron along the z-direction. Given that the oscillation of the in-plane magnetic moment around zero is relatively small compared to the out-of-plane magnetic moment, the transfer of spin charge is only considered in the z direction (still described using spin-up and spin-down). The timedependent charge dynamics ($\Delta$n (t)) is depicted in Figure 2b, which illustrated that the reduction in the Cr$_1$ atomic magnetic moment is mainly attributed to the flipping of spin electrons in Cr$_1$ atom, which is the reduction in spin-up electrons and the increase in spin-down electrons. Additionally, there are part charge transfer from Br and S atoms. 

In order to facilitate a comprehensive comprehension of the mechanism of electron movement, Figure 2c was devised. To summarize, during the initial 18 fs of excitation, the spin flip processes involved are as follows: Br\underline{~}up, Cr\underline{~}up to Cr\underline{~}down, S\underline{~}down to S\underline{~}up, and the spin transfer processes include: Br\underline{~}down, S\underline{~}down to Cr\underline{~}down, Cr\underline{~}up to Cr\underline{~}up. Subsequent to 18 fs, electron movement is predominantly characterized by a transition from a high-energy excited state to a low-energy excited state. The vast majority of excited state electrons relax to the Cr\underline{~}up band of the CBM. While the alterations in magnetic moment that under static conditions are a consequence of the transfer of electrons between the same spin-states or transfer to the interstitial region.

\begin{figure*}[t!]
	\centering
	\includegraphics[width=0.9\textwidth]{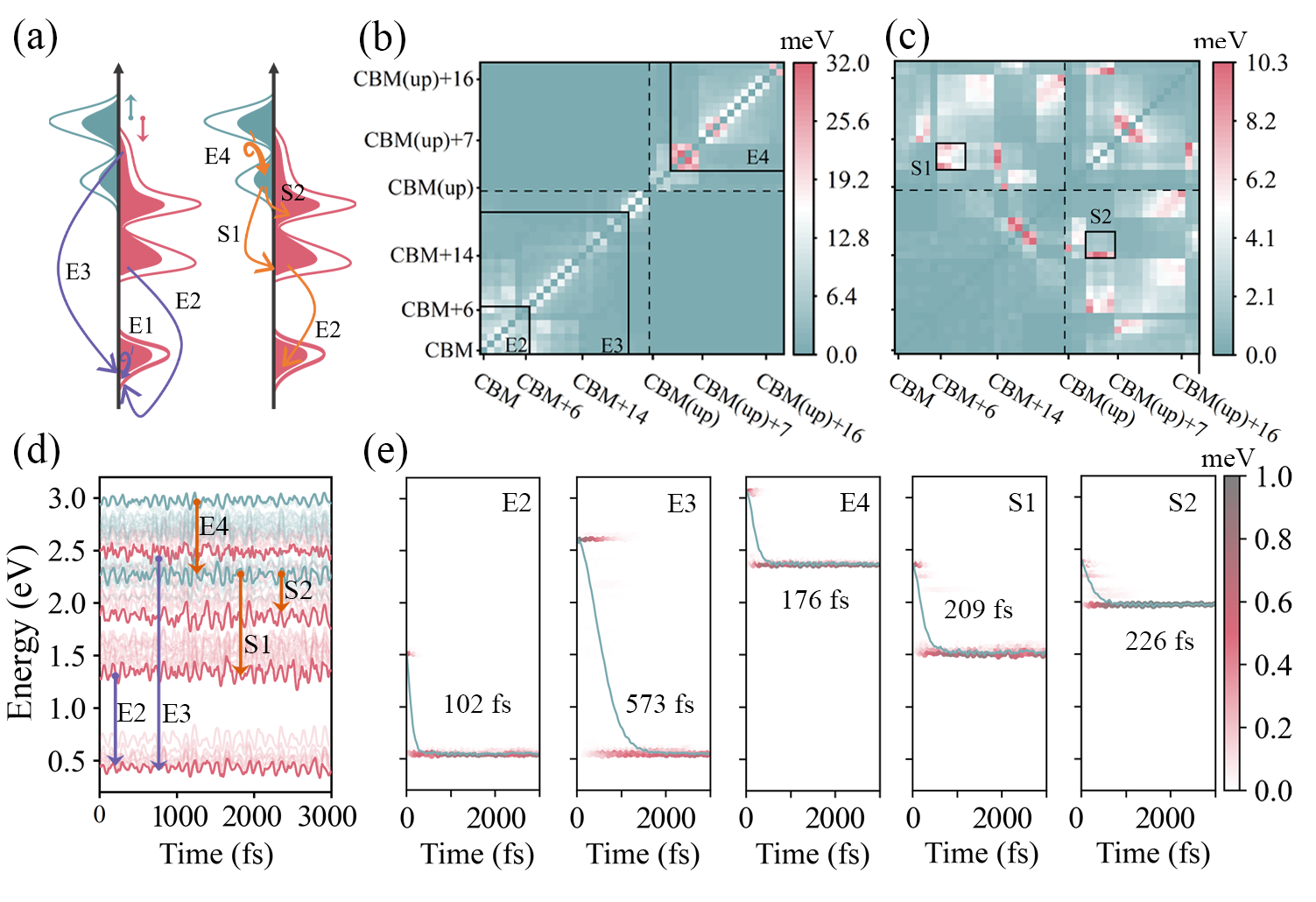} 
	\caption{NAMD simulation of spin relaxation dynamics in considering all-phonons mode for CrSBr monolayer. (a) Schematic diagram of spin relaxation. E1-E4 represent the spin-transfer processes controlled by EPC, S1 and S2 represent the spin-flip processes controlled by SOC. (b) and (c) EPC and SOC components between key electronic states, frames represent the corresponding NAC component of different processes. (d) Time evolution of the spin-resolved energy states at the $\Gamma$ point, the blue, pink lines are represented spin-up and spin-down spin states. The purple and orange arrows are indicated the relaxation process of spin-up and spin-down electrons, respectively. (e) Energy relaxation of the excited electrons. The color map indicates orbital localization.}
\end{figure*}

To elucidate the role of lattice degree of freedom in demagnetization process, we examine the CrSBr relaxation dynamics under the spin-phonon coupling system, shown in Figure 2d-f. As previously stated, the stage \Rmnum{1} of demagnetization is in accordance with the static nuclei and static nuclei+SOC system. It is crucial to note that the phonon system does not exhibit the thermal balance phase. The distribution of larger EPC is primarily confined to adjacent band components with same spin orientation. In the CrSBr monolayer, spin-down electrons demonstrate faster relaxation to the CBM than spin-up electrons. The reason for this is that, due to the occupation of the CBM by spin-down electrons, the relaxation process from the higher excited state to the CBM is more straightforward than that of spin-up electrons from the spin-up CBM (which is referred to as CBM(up)) to the CBM, which is controlled by SOC. Consequently, the persistent decline in the magnetic moment following 18 fs can be attributed to the uninterrupted relaxation of the same spin state electronics (Figure 1c).

As with SOC, an explanation is provided for the movement of electrons in the phonon system through changes in spin charge (see figure 2d-f). When phonons are considered in the system, the SOC effect is found to be negligible, as evidenced by the minute change in the in-plane magnetic moment depicted in Figure 2d. The pulse excitation process in stage \Rmnum{1} is characterized by a series of charge transfers, involving Br\underline{~}up, Cr\underline{~}up to Cr\underline{~}up, Br\underline{~}down, S\underline{~}down to Cr\underline{~}down, and Cr\underline{~}up to S\underline{~}up. This phenomenon is attributed to the weak SOC effect, which still involves the transfer of some electrons to the interstitial region. As time progresses, the interaction between the electronic system and the phonon system gradually intensifies, resulting in the effective transfer of electron energy to the phonon system. Concurrently, the lattice thermalization (i.e., the increase in phonon energy) will precipitate thermal expansion and enhanced lattice vibration, which will ultimately be dissipated through heat conduction or radiation. This phenomenon is also the underlying cause of the dissipation of electrons in each atom in the subsequent stage, whereby a proportion of electrons relax to the CBM spin-down energy band, while a separate proportion of them transfer to the interstitial area. Phonons play a pivotal role in the processes under discussion; consequently, a comprehensive study will be conducted to explore the impact of diverse vibrational modes of phonons on electron relaxation.

So far, it has been proven that the excitation of electrons by the pulse occurs in the first 20 fs, resulting in the generation of hot electrons, and that the effects of SOC and EPC on the spin dynamics after 20 fs are crucial. The present study employed first-principles nonadiabatic molecular dynamics (NAMD) simulations to investigate the relaxation of hot carriers when considering SOC and EPC to discuss the root cause of the second demagnetization. The present study simulated the relaxation process of electrons from a high-energy excited state to the final state (CBM) when the electrons interacted with all-phonon or a sigle specific phonon mode. The all-phonons vibration mode is utilised to simulate the real electron relaxation process, while the single-phonon is employed to investigate the effect of the coupling between a specific vibration mode and high-energy electrons on the relaxation process.

When all-phonons mode are considered, the CBM is occupied by spin-down electrons, and the energy band shows that the spin-up CBM(up) is 1.58 eV higer than CBM. Therefore, the relaxation of spin-down electrons to the CBM is facilitated by EPC (process E1, E2,or E3), while the relaxation process of high-energy spin-up electrons is more complicated and includes three processes: relaxation to the vicinity of CBM(up) through EPC (process E4). Subsequently, these electrons flip to the spin-down energy band through SOC (process S1 or S2), and finally repeat the processes E1, E2 and E3. The schematic representation of the spin-down and spin-up electron relaxation process is depicted in Figure 3 (a). The conclusion drawn from this analysis is that the decline in spin-up electrons with the rise in spin-down electrons, results in a reduction of the magnetic moment. Next, the non-diabatic coupling (NAC) matrix of key electronic states in CrSBr was calculated, thus facilitating the clear determination of the contribution of EPC and SOC components (see Figure 3b, 3c). The research findings indicate a correlation between EPC intensity and the transition of electrons between adjacent energy bands, particularly within the same spin state \cite{PhysRevB.105.085142}. This transition involves the transfer of energy from the electrons to the phonons through electron-phonon scattering. The coupling between different spin states and specific energy bands in the SOC is of particular importance, especially in the transition between high energy bands and low energy bands with differernt spin-state.

As illustrated in Figures 3d and 3e, the energy states implicated in the relaxation process of the spin-up and spin-down excited states within the energy window [0.2, 3.2] eV are demonstrated, along with the respective relaxation times of each process. The decay time scale, $\tau$, of each relaxation process is fitted using a Gaussian function \( f(t) = \emph{\text{a}} + \emph{\text{b}} \exp (-\emph t / \tau) \). If the in the electrons of valence band are excited to orbitals at 0.68, 1.29, and 2.50 eV, respectively, which are the same spin states as the CBM. They will relax to CBM after the EPC driver in 199, 102 and 573 fs, respectively. These processes are designated as process E1, E2, and E3, respectively. It can be found that the relaxation time of electronics is predominantly influenced by the following factors: 1. The energy difference between bands, and as the value smaller, the relaxation time decreases concomitantly (E2 compare to E3); 2. EPC interaction, a large energy spacing can also facilitate electron interaction with different sounds during the relaxation process, thereby enabling rapid energy release (E1 compare to E2).

In a similar vein, the relaxation process of spin-up electrons to CBM was examined, and it was hypothesized that the SOC component is the primary controlling factor, which will begin to relax from the 2.03 (CBM(up)) and 2.26 eV (CBM(up)+3) energy bands.The relaxation times obtained by the fitting process are 1308 and 232 fs, respectively. It was observed that the relaxation time for electrons to transition from the CBM(up)+3 energy band to CBM is less than that for CBM(up).The calculation of the SOC component NAC in Figure 3c indicates that the coupling strength between the degenerate energy bands CBM(up)+3 and above with the spin-down energy level is stronger. This suggests that an increase in NAC will result in a decrease in relaxation time. Furthermore, the relaxation time from the 2.26 eV spin-up energy band to the CBM is shorter than the duration of the E3 process. However, the SOC coefficient is considerably smaller than EPC, indicating that the relaxation process of spin-up electrons necessitates a flipping to the spin-down state and subsequent EPC relaxation. To this end, the time of two strong SOC state flipping processes (processes S1 and S2) was measured to be 209 fs and 226 fs, respectively, which can verify the above speculation.

During material excitation by ultrafast pulses, electron transitions generate vibrations that produce phonons, enabling non-radiative energy transfer to the lattice. CrSBr exhibits nine Raman-active vibrational modes, and to clarify their roles in electron relaxation, we analyzed five key pathways through single-phonon dynamics. These studies reveal that specific vibrational modes govern distinct relaxation channels, with each phonon-electron interaction exhibiting a characteristic simplicity. Single-phonon-dominated processes demonstrate how targeted vibrational excitations mediate energy dissipation, highlighting mode-specific contributions to spin-lattice coupling.

\begin{table}[htp]
	\renewcommand{\arraystretch}{1}
	\caption{The relaxation time (fs) of electrons in single-phonon and all-phonons mode through E2, E3, E4, S1 and S2 processes. The "$\times$" indicates that the electron relaxation process is blocked and is unable to reach the CBM.}
	\label{tab1}
	\centering
	\begin{tabular}{ccccccccccc}
		\hline
		{~}& A$_g^1$& A$_g^2$& A$_g^3$& B$_{2g}^1$& B$_{2g}^2$& B$_{2g}^3$& B$_{3g}^1$& B$_{3g}^2$& B$_{3g}^3$& All\\
		\hline
		E2& 314& 215& 166& 229& 108& 110& \textbf{83}& 159& 120& 102\\
		E3& $\times$& $\times$& $\times$& 468& 1545& $\times$&$\times$& $\times$& $\times$& 573\\
		E4& 599& 750& 454& 687& 877& 396& 3345& 10995& 11993& 176\\
		S1& $\times$& 441& $\times$& 644& 1641& 553& 785& 909& 913& 209\\
		S2& 357& 497& 712& 855& 1913& 389& 931& 454& \textbf{324}& 226\\
		\hline
	\end{tabular}
\end{table}

\begin{figure*}[t!]
	\centering
	\includegraphics[width= 0.8\textwidth]{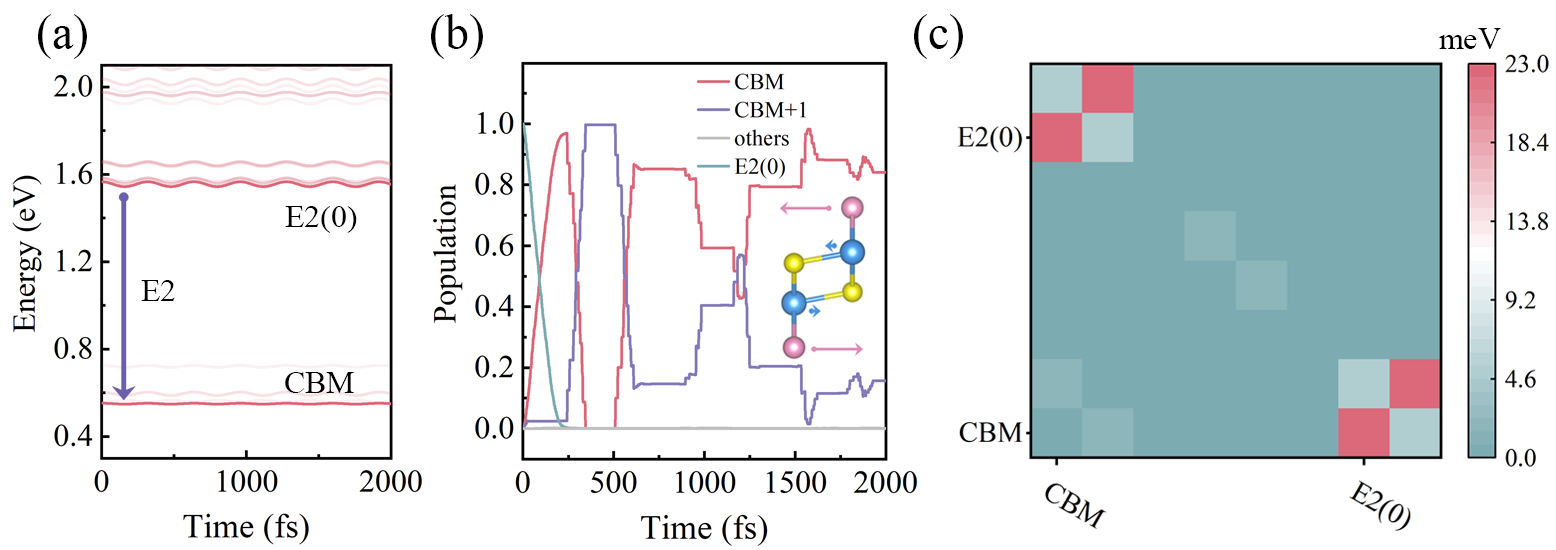} 
	\caption{ NAMD simulation of spin relaxation dynamics in considering the single-phonon mode of B$_{3g}^1$ phonon for CrSBr monolayer. (a)Time evolution of the spin-resolved energy states at the $\Gamma$ point. The spin-down energy band E2(0) (initial energy band of E2 processs) to CBM is associated with the relaxation process E2. (b) Time evolution of the spatial distribution of excited-state electrons population on processes E2. Insets illustrate atomic displacements corresponding to B$_{3g}^1$ vibration mode, the arrows designated as positive. (c) EPC components between key electronic states, frames represent the corresponding NAC component of E2 process.}
\end{figure*}

Table 1 records the relaxation time of electrons in single-phonon and all-phonons mode through E2, E3, E4, S1 and S2 processes. It has been established that the relaxation process of electrons from the 1.29 eV spin-down state to the CBM (process E2) is shortened to 83 fs when there is only a single B$_{3g}^1$ phonon vibration mode, and the remaining phonon modes all delay the relaxation time to a certain degree compare with all-phonons mode. This finding suggests that phonon modes with B$_{3g}^1$ frequency better match electron relaxation energy levels, thereby facilitating more direct and efficient energy transfer. However, when electrons transition from higher energy levels to lower energy levels (process E3 or E4), multi-phonon cooperative interactions are typically implicated.Consequently, if single-phonon relaxation channels are the sole consideration, relaxation blockade ensues, signifying that electrons are incapable of relaxing to the CBM or the relaxation time is augmented. A significant number of single-phonon channels are blocked when electrons undergo the E3 process, and only the B$_{2g}^1$ and B$_{2g}^2$ phonon channels are conducive to the process.

The spin-phonon couplings is achieved through the vibration modes of phonons in the lattice. The effect of these changes on the orbital motion and spin direction of electrons is such that the process is complicated. The relaxation of spin-up electrons is comprised of two distinct processes: spin flip and spin transfer. This necessitates the involvement of a non-radiative process assisted by multiple phonons. From the perspective of the E4 process, any of the single-phonon vibration mode will increase the relaxation time of the electron, among which the most significant is the B$_{3g}$ phonon. It is noteworthy that a similar phenomenon is observed in the spin flip process (process S1), which also exhibits a shorter relaxation time when multiple phonons are involved. However, it has been ascertained that the relaxation time of the single phonon vibration modes B$_{3g}^3$ and A$_{1g}$ in the S2 process is relatively proximate to that of the full phonon vibration mode. It can be deduced that these two phonon vibration modes play a pivotal role in the electron spin flip S2 process. In concluded that the alteration in the symmetry of the local atomic environment, occasioned by the relevant phonon vibration mode, will result in a modification of the coupling strength between spins and orbitals. This, in turn, will lead to a change in the electron relaxation time that will meet the requirements for practical application.

It has been ascertained that the CBM+1 state in the E2 process of the B$_{3g}^1$ phonon mode is occupied by electrons alternately with the CBM during the relaxation process (Figure 4b). The underlying reason can be found in the NAC coupling matrix (Figure 4c). The coupling strength between E2(0)+1 and CBM+1 is equivalent to that between E2(0) and CBM, indicating that the probability of electron transition between these two states is equivalent. In the B$_{3g}^3$ phonon mode, the S2 process, governed by SOC, results in electrons occupying only two specific electronic energy states, the spin-up S2(0) (initial energy band of S2 process) to spin-down S2(1) (final energy band of S2 process) (Figure S3b). The NAC matrix representing SOC (Figure S3c) demonstrates that the coupling strength between CBM(up)+3 and CBM is considerably stronger than that of other energy bands, indicating that electrons will directly transition between these two energy states without occupying other energy bands. Additionally, it was observed that phonons in plane vibration modes interact more readily with electrons in two-dimensional materials, leading to scattering and transfer of momentum and energy.

\begin{figure*}[t!]
\centering
\includegraphics[width= 0.8\textwidth]{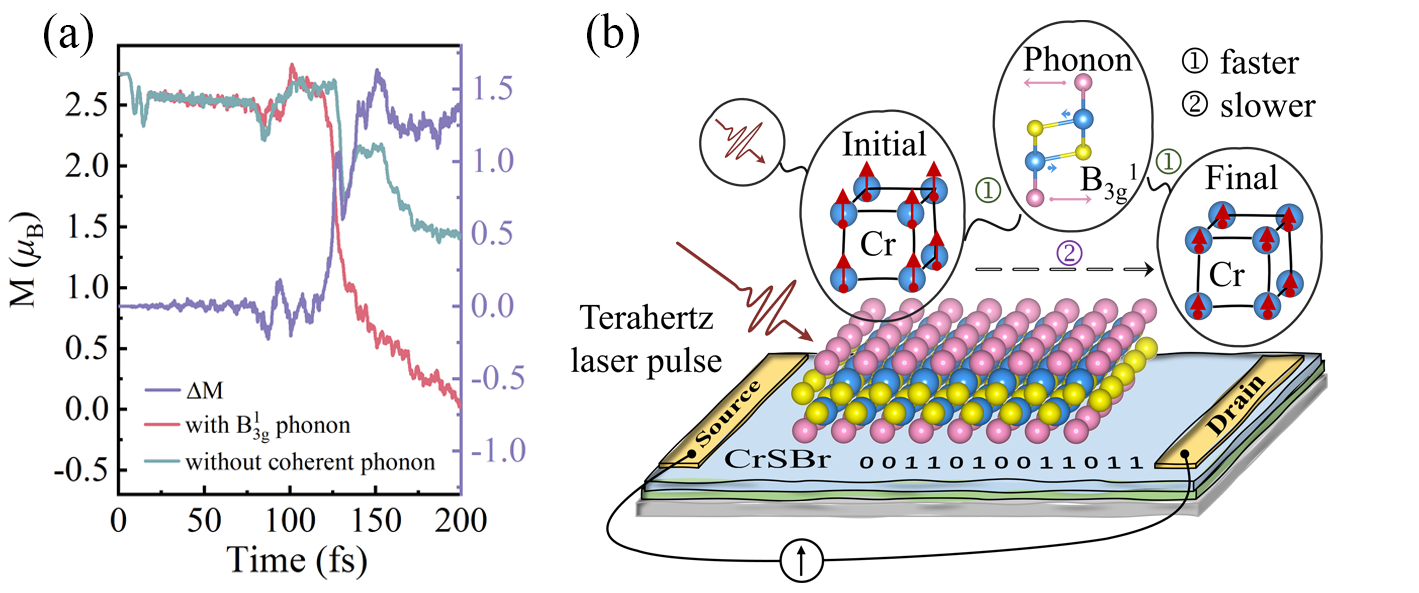} 
\caption{ (a)Time-dependent magnetic moment dynamics of Cr$_1$ atom contrast of the coherent phonon modes (B$_{3g}^1$) in the absence of nuclear dynamics. The $\Delta$M is defined as the difference in demagnetization between the B$_{3g}^1$ coherent phonon mode and without coherent phonon mode. (b) Schematic diagram of coherent phonon-controlled information storage after pre-excitation of a monolayer CrSBr by a terahertz femtosecond laser pulse.}
\end{figure*}

The realization of the dual-pumping method makes it possible to study the spin dynamics of coherent phonons. During the dual-pumping method, the lattice displacement generated by the first laser beam is called a coherent phonon \cite{PhysRevLett.73.740, sun2023coherent}, and then the second laser beam is used to manipulate the material spin dynamics of the existing excited phonons. We analyzed the coherent demagnetization in the above-mentioned B$_{3g}^1$ phonon vibration mode. As shown in Figure 5(a), the demagnetization of Cr atoms exhibits similar oscillation behavior. Compared with the absence of coherent phonons, the effect of precoherent phonons on the demagnetization amplitude is as high as about 1.5 $\mu $$_B$ at 200 fs. This is because the lattice vibration mode induced by coherent phonons enhances the spin-orbit intensity, thereby accelerating the demagnetization process, indicating that the pre-excited coherent phonons provide a more efficient energy transfer pathway, allowing the energy of the electronic system to be dissipated into the lattice after laser excitation. That is, when the THz laser pulse irradiates the monolayer CrSBr, the electrons inside the material interact with the lattice, inducing a specific lattice vibration mode, namely, coherent phonons. This phonon mode can further modulate the spin dynamics of the material during optical excitation, thereby affecting its application in high-speed information storage and processing.

The mechanism and application prospects of terahertz laser pulse-induced magnetization dynamics in monolayer semiconductor CrSBr are illustrated in Figure 5(b). The system loses equilibrium under the action of the laser, triggering a transient lattice response characterised by a phonon mode. This mode functions as an ultrafast channel (marked as path \textcircled{1}), facilitating the rapid alignment of the Cr magnetic moment towards the ferromagnetic final state (approximately 0 $\mu $$_B$). In contrast, the slower relaxation process (path \textcircled{2}) also leads to a demagnetization process, but the demagnetization amount is relatively small when the spin-lattice coupling effect is ignored. The detection of the encoded spin information in the Cr magnetic order is subsequently facilitated by the source-drain electrodes, thereby demonstrating the prospect for light-driven magnetic memory or spin logic applications.

The discovery of the all-optical magnetization switching of CrI$_3$ shows the application of magnetic control in controlling low-power and high-speed spin electronic devices \cite{zhang2022all}, and also inspires theoretical research on light-induced ultrafast processes. Zhou et al. and Wu et al. reported in Fe$_3$GeTe$_2$ (FGT) monolayers/bulks, the important role of electroacoustic coupling in spin dynamics after ultrashort laser pulses, and a third-order demagnetization process has been proposed \cite{zhou2023ultrafast, wu2024three}. They directly used the phonon vibration modes with strong Raman activity (basically out-of-plane phonon vibration modes) as coherent phonons to study light-induced spin dynamics. Our research first explored the non-adiabatic electron relaxation dynamics of all Raman vibration modes, obtained the relative electron relaxation time, and then studied the effect of coherent phonons on the demagnetization process. This method can comprehensively and accurately obtain the pre-excited B$_{3g}^1$ phonon vibration mode of the monolayer CrSBr material, which can greatly accelerate the light-induced demagnetization process and increase the amount. Moreover, this mode is in-plane vibration, and the electron relaxation occurs within 83 fs (emphasizing the influence of phonons on spin dynamics within 100 fs). After in-depth research, it was found that the relaxation process of spin electrons can be precisely controlled by controlling the pre-excited phonon state. Our research further expands the understanding of spin-lattice coupling in two-dimensional magnetic materials and provides guidance for experimental research.

\section{CONCLUSIONS}
In conclusion, this study provides a comprehensive understanding of the ultrafast photoinduced demagnetization dynamics in monolayer CrSBr, highlighting the intricate interplay between spin, electron and phonon degrees of freedom. Our results reveal a two-stage demagnetization process: an initial rapid electron-driven demagnetization, followed by a longer-lasting phase in which spin is coupled to light-induced lattice vibrations. We show that spin-lattice coupling plays a crucial role in the second stage, with electron-phonon interactions significantly influencing the overall dynamics. We also reveal the important role of different phonon vibrational modes in the process of electron relaxation dynamics. Furthermore, experimental evidence has demonstrated that the impact of the B$_{3g}^1$ phonon vibration mode on the electron relaxation of the CrSBr monolayer is within 83 fs. The results not only deepen our understanding of material behavior under intense laser pulses, but also open new avenues for controlling spin dynamics in 2D materials for applications in next-generation electronic and quantum systems.

\section*{ACKNOWLEDGEMENTS}
This work is supported by National Natural Science Foundation of China (No.12374057, 12204069), the Startup funds of Outstanding Talents of UESTC (A1098531023601205), Sichuan Science and Technology Program (Grant No. 2024NSFSC1385) and Fundamental Research Funds for the Central Universities. L. Y acknowledges the Natural Science Foundation of Hunan Province (Grant No. 2025JJ60060). The work (S.T.) at Los Alamos National Laboratory (LANL) was performed in part at the Center for Integrated Nanotechnologies (CINT), a U.S. Department of Energy, Office of Science user facility at LANL.

\bibliographystyle{apsrev4-2}
\bibliography{CrSBr}
\end{document}